\documentclass[twocolumn,showpacs,preprintnumbers,groupedaddress,amsmath,amssymb,floatfix,prl]{revtex4}
\usepackage{graphicx}
\usepackage{epsfig}
\usepackage{ulem}
\usepackage{setspace}
\begin{document}

\title{Observation of an ultracold plasma instability}

\author{X. L. Zhang, R. S. Fletcher, and S. L. Rolston}

\affiliation{Joint Quantum Institute and Department of Physics, University of Maryland, College Park, MD 20742}
\date{\today}

\begin{abstract}
 We present the first observation of an instability in an expanding ultracold plasma. We observe periodic emission of electrons from an ultracold plasma in weak, crossed magnetic and electric fields, and a strongly perturbed electron density distribution in electron time-of-flight projection images. We identify this instability as a high-frequency electron drift instability due to the coupling between the electron drift wave and electron cyclotron harmonic, which has large wavenumbers corresponding to wavelengths close to the electron gyroradius.
\end{abstract}


\maketitle

Ultracold plasmas (UCPs), produced by photoionizing a sample of laser-cooled and trapped atoms, have extended neutral plasma parameters by over two orders of magnitude, to electron temperatures below 1 K \cite{killian1999}. Studies of UCPs have primarily concentrated on temperature measurements \cite{roberts2004, gupta2007, fletcher2007}, and expansion dynamics \cite{kulin2000, bergeson2003, robicheaux2003}, and recent work has identified a stable collective mode \cite{fletcher2006}. A signature of the collective and nonlinear nature of plasmas is the existence of plasma instabilities, perturbations that grow exponentially to large amplitudes and dominate plasma dynamics. Much of the quest for fusion energy involves control and suppression of plasma instabilities \cite{conway2000}. This universal dynamics occurs in all kinds of plasmas, including space plasmas \cite{ oppenheim2003, farley1963, buneman1963}, dusty plasmas \cite{Merlino1998}, magnetically confined plasmas \cite{krall1971}, and Hall thruster plasmas for spacecraft propulsion \cite{litvak2004, lazurenko2005, adam2004, ducrocq2006}. 
  
 In this work, we present the first observation of a plasma instability in an expanding UCP. By applying a small magnetic field ($\sim$ 2 G) perpendicular to an applied electric field ($\sim$20 mV/cm), we observe periodic pulsed emission of electrons from an expanding UCP, with a frequency range from 50 to 500 KHz. Using a time-of-flight electron projection imaging technique \cite{zhang2008}, we image the electron spatial distribution by extracting them with a high-voltage pulse and accelerating them onto a position-sensitive detector. We observe that electron projection images split into two or three lobes in the $E \times B$ direction, coincident with the observation of periodic electron emission signals. This provides strong evidence for a plasma instability in the expanding UCP due to the electrons drifting relative to the ions across the magnetic field. A high-frequency electron drift instability \cite{adam2004, ducrocq2006} quantitatively matches our observation, which has a frequency lower than the electron gyrofrequency and a short wavelength on the order of the electron gyroradius, due to the coupling between the electron drift wave and a harmonic of the electron cyclotron frequency. 

Our production of UCPs is similar to our previous work \cite{killian1999}, which we briefly summarize. We cool and trap about a few million metastable Xenon atoms in a magneto-optical trap (MOT). The neutral atom cloud has a temperature of about 20 $\mu$K, peak density of about 2 x 10$^{10}$ cm$^{-3}$, and Gaussian spatial density distribution with an rms radius of about 300 $\mu$m. We then produce the plasma by a two-photon excitation process, ionizing up to 30$\%$ of the MOT population. One photon for this process is from the cooling laser at 882 nm, and the other is from a pulsed dye laser at 514 nm (10-ns pulse). We control the ionization fraction with the intensity of the photoionization laser, while the initial electron temperature $T_{e0}$ (typically around 1-500 K) is controlled by tuning the 514-nm photon energy with respect to the ionization limit. 

\begin{figure}[htbp]
\begin{center}
\epsfig{file=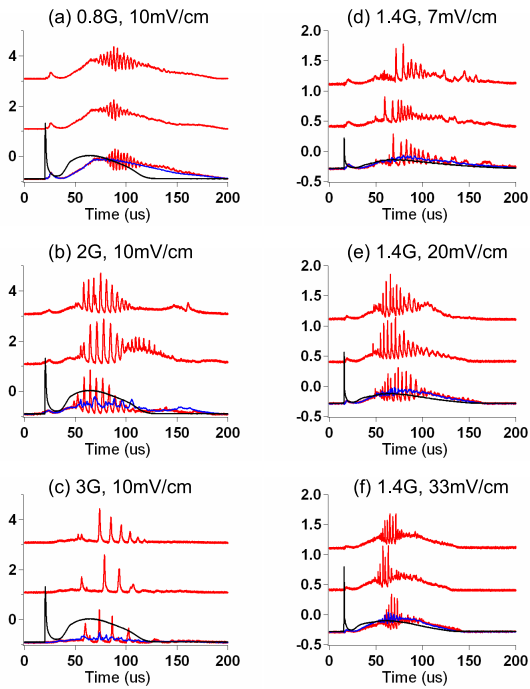, width=3.0in}
\end{center}
\caption{ Electron emission signals for different E and $B_\perp$. (a)-(c) constant E and different $B_\perp$; (d)-(f) constant $B_\perp$ and different E. The red curves (with large periodic emission) are single shot; the blue curve (with much less periodic emission) is the average of 40 shots. the black curve (with large prompt peak) is the electron signal without a magnetic field.}
\label{figure1}
\end{figure}

The ionized cloud rapidly loses a few percent of the electrons, resulting in a slightly attractive potential for the remaining electrons, and quickly reaches a quasineutral plasma state. It then expands with an asymptotic velocity (50-100 m/s) caused by outward electron pressure \cite{kulin2000}, and maintains a roughly Gaussian density profile during the lifetime of the UCP \cite{zhang2007}. There are two grids about 1.5 cm above and below the plasma to provide a small electric field E ($\sim$5-50 mV/cm) so that electrons leaving the plasma are guided to a microchannel plate detector. We can apply a longitudinal magnetic field $B_\parallel$ parallel to E, or a transverse magnetic field $B_\perp$ perpendicular to E. The applied magnetic fields are turned on before the two-photon excitation process. 
The black curve in Fig.~\ref{figure1} is the typical electron emission signal from a freely expanding UCP for $T_{e0}=100$ K. The signal consists of a prompt peak (initially escaped electrons) followed by a region of little electron loss where the quasineutral plasma state forms. This is followed by a long $\sim$150 $\mu$s loss of electrons, interpreted as the decay of the plasma as electrons spill out of the potential well, which gets shallower as the plasma expands. The electron signal with a small longitudinal $B_\parallel$ ($\sim 12$ G) looks similar to that without a magnetic field except for small enhancement and some changes in the expansion dynamics (expansion dynamics in a large $B_\parallel$ is studied by time-of-flight projection imaging technique \cite{zhang2007}). 

The electron emission signal with a small transverse magnetic field $B_\perp$ is dramatically different from that with a small $B_\parallel$. We observe periodic pulsed electron emission even with a very small $B_\perp$, as shown in Fig.~\ref{figure1}a-\ref{figure1}c. As we increase $B_\perp$ from zero, the electron signal starts to have oscillations at about 0.8 G. The periodic emission appears at about 30-50 $\mu$s after the formation of the plasma with frequency of several hundred KHz. The three traces in each panel of Fig.~\ref{figure1} correspond to individual single realizations of the UCP. Note that the emissions have similiar character for each shot, although the phases are random. As we continue to increase $B_\perp$ to about 2-2.5 G (Fig.~\ref{figure1}b), the amplitude of the emissions gets larger, comparable to the prompt peak in the absence of a magnetic field. The frequency decreases to 50-100 KHz and the prompt peak gets even smaller. The background electron emission signal almost vanishes except for few peaks at about 3-3.5 G (Fig.~\ref{figure1}c).
 
\begin{figure}[htbp]
\begin{center}
\epsfig{file=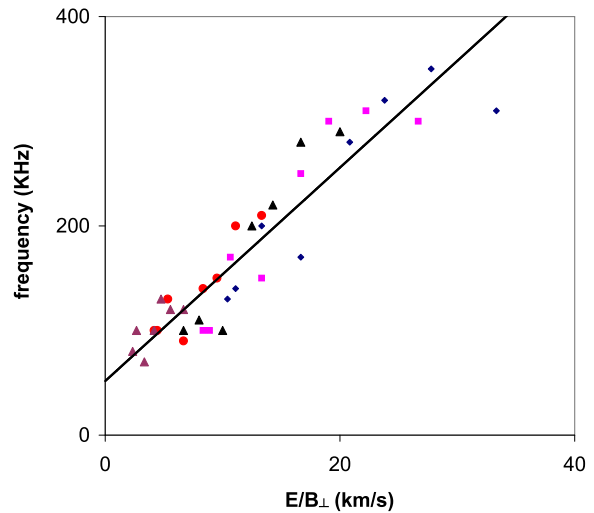, width=3.0in}
\end{center}
\caption{ The emission frequency as a function of $E/B_\perp$; each symbol for varying $B_\perp$ from 1 to 3.2 G and constant E; different symbol for varying E from 7 to 33 mV/cm and constant $B_\perp$. The emission frequency is linearly dependent on the electron drift velocity $V_d$.}
\label{figure2}
\end{figure}

The frequency and amplitude of the periodic electron signal also depends on the applied electric field E (Fig.~\ref{figure1}d-\ref{figure1}f). The frequency increases and the amplitude of the emissions decreases as we increase E, which is similar to the case of decreasing $B_\perp$. Using fast Fourier transform (FFT) of the electron signals in Fig.~\ref{figure1}, we can extract the emission frequency as a function of $E/B_\perp$ (electron drift velocity $V_d$), as shown in Fig.~\ref{figure2}. The emission frequency depends linearly on the drift velocity with a slope of  $10.2\pm0.2$ $m^{-1}$.
      
\begin{figure}[htbp]
\begin{center}
\epsfig{file=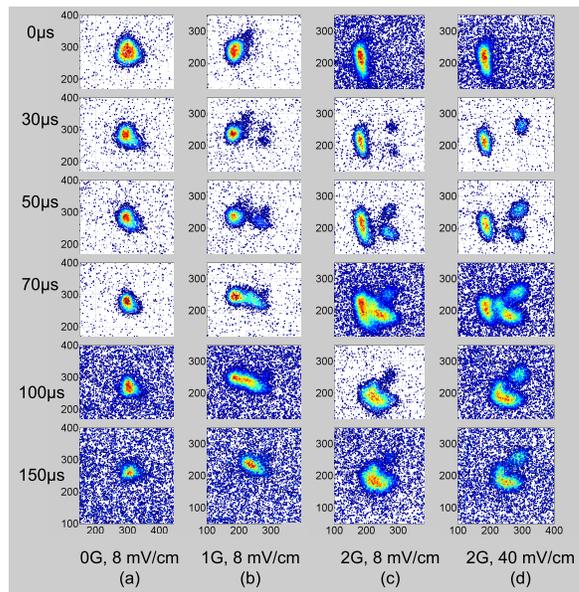, width=3.0in}
\end{center}
\caption{ Electron projection images in weak crossed magnetic field and electric field. All the units of the images here are in pixels, and one pixel is about 150 $\mu$m. }
\label{figure3}
\end{figure}

By using a time-of-flight projection imaging technique \cite{zhang2008}, we image the electron spatial distribution onto a position-sensitive detector (a phosphor screen attached to a multichannel plate detector), and record images with a CCD camera. Figure~\ref{figure3} shows the electron projection images at different delay times after the formation of the UCP for different $B_\perp$ and E. All the images are single shot, with little variation from shot to shot. In the absence of a magnetic field (Fig.~\ref{figure3}a), the electron images show a Gaussian density profile. Note that the electron image size slowly decreases during the lifetime of the UCP because of the strong Coulomb force of the ion cloud on the electrons, electron loss and electron Coulomb explosion effects, unlike ion images (whose sizes decrease in the first 20 $\mu$s due to the strong Coulumb explosion effect of the dense ion cloud, and then increase afterward, reflecting the real size of the plasma) \cite{zhang2008}. As we increase $B_\perp$ to about 1 G (Fig.~\ref{figure3}b), the electron images start to split into two lobes at about 30 $\mu$s, coincident with the observation of large periodic electron emission signals. At about 2 G, we observe up to 3 lobes in the electron images between 30 $\mu$s and 70 $\mu$s (Fig.~\ref{figure3}a and \ref{figure3}d). The extra electron lobes are in the E$\times B_\perp$ direction, and they show up in the other side of the main electron cloud if we change the sign of $B_\perp$. We did not observe any changes in the ion images and ion current in the cross-field configuration compared to those without a magnetic field. 
This is quite surprising, as one typically expects the electron distribution to closely follow the ion distribution for a cold neutral plasma.
In order to confirm that the extra lobes of the electron images are coming from the plasma itself, rather than dynamics during the flight to the detector of the projection imaging technique, we took a series of images at different extraction parameters such as the high-voltage amplitude and width and the accelerating voltages on the middle and front grids (located between the plasma and the detector, and strongly affect the sizes of electron images). We always observe the similiar results as Fig.~\ref{figure3}, implying this is a good measure of the electron distribution in the plasma. Using a simple model of the electron space charge effects during transit to the detector, we find the separation between the main cloud and the extra lobes before time of flight to be about 0.5-1$\sigma_t$, where $\sigma_t$ is the plasma size at delay time t. That is, the extra lobes are located inside the plasma.


There are a multitude of different instabilities in plasmas, but given our parameters we can constrain possible choices to a small number. We identify the periodic electron emission signals as well as the extra lobes in the electron projection images as a signature of a plasma instability due to electrons drifting relative to ions across the magnetic field. The crossed magnetic and electric fields will drive the electrons to drift with a velocity $V_d$ ($V_d = E/B_\perp$) in the $E \times B_\perp$ direction. The ions are unmagnetized, not affected by the small $B_\perp$ (several G), due to their large mass (the ion gyroradius is much larger than the UCP size and the ion gyro-period is much longer than the UCP lifetime). In our UCP the electrons are unmagnetized in the first $\sim$30 $\mu$s because the electron gyroradius is about the same order of magnitude as the UCP size and the electron collision rate is higher than the electron cyclotron frequency. As the UCP expands, the plasma size gets larger, the plasma density gets lower and the electron temperature gets smaller (due to various cooling mechanisms) \cite{fletcher2007}. The electrons become magnetized at about 30-50 $\mu$s after the formation of the UCP because the electron collsion rate starts to be less than the electron gyrofrequency and the plasma size at that time is about a factor of 10 larger than the electron gyroradius, which is consistent with our observation that the periodic emissions start at the same time. 
The frequency is in the range from 50 KHz up to a few hundred KHz, which is much larger than the ion cyclotron frequency $f_{ci}$ ($\sim$12 Hz at 1 G) and much less than the electron cyclotron frequency $f_{ce}$ ($\sim$2.8 MHz at 1 G) and the electron plasma wave $f_{pe}$ which is about 10-20 MHz at the delay time of about 30-50 $\mu$s. The periodic emission signal is roughly independent of plasma density (or plasma frequency) as shown in Fig.~\ref{figure1} (note that the plasma density drops by a factor of 8 from 50 $\mu$s to 100 $\mu$s). There are periodic emissions with different frequencies (Fig.~\ref{figure1}b and \ref{figure1}d), which also indicates that we have plasma instability (mode switching) in the cross-field configuration.

A candidate instability is the high-frequency electron drift instability, which has been studied in Hall thrusters (a type of plasma-based propulsion systems for spacecrafts) experimentally \cite{litvak2004, lazurenko2005} and theoretically \cite{adam2004, ducrocq2006} in the past few years to explain the transport of electrons across the magnetic field lines. A 2D fully kinetic model of the Hall thruser developed in ref.~\cite{adam2004} has demonstated that the large drift velocity at the exhaust of the thruster was the source of an instability that gives rise to plasma turbulence and could induce a significant current across the magnetic field. It is a high-frequency electron drift instability with frequency $f_r$ lower than the electron cyclotron frequency $f_{ce}$ and short wavelength close to the electron gyroradius $r_{ce}$, which is studied in ref.~\cite{ducrocq2006}. 

The theory of high-frequency electron drift instability is developed from the dispersion equation of electrostatic waves in a hot magnetized electron beam drifting across a magnetic field with unmagnetized cold ions, and is closely related to the ion-acoustic-wave instabilities in other plasma systems, such as non-specular radar meteor trails \cite{oppenheim2003}, the ionosphere of the earth \cite{farley1963, buneman1963}, dusty plasmas \cite{Merlino1998} and magnetic pulses \cite{krall1971}, but they usually restrict the analysis to the cases where the drift velocity $V_d$ is much smaller than the electron thermal velocity $V_{eth}$. For high-frequency electron drift instabilities in Hall thrusters and here, $V_d$ is about the same order of magnitude as the $V_{eth}$, and much larger than the magnetic field gradient drift velocity and the density gradient drift velocity. 
 
\begin{figure}[htbp]
\begin{center}
\epsfig{file=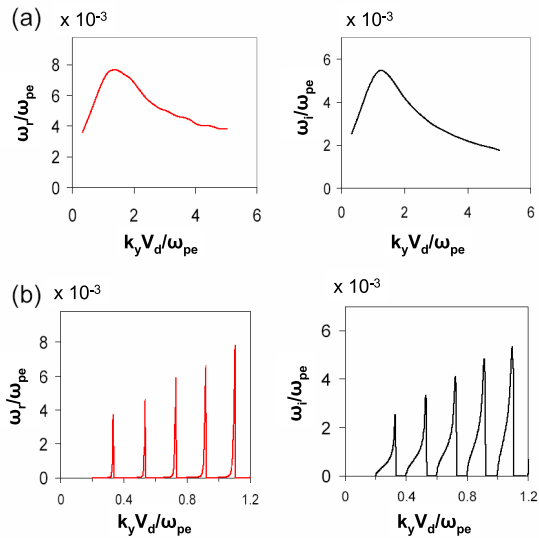, width=3.0in}
\end{center}
\caption{ Numerical solutions of the 1D dispersion relation of equation (1) as in \cite{ducrocq2006}, but for our condition. (a) represents the envelopes of the real (frequency) and the imaginary (growth rate) as a function of $k_y V_d/\omega_{pe}$. (b) the corresponding ones for small wavenumbers. The electron cyclotron frequency is equal to $0.2 \omega_{pe}$, and $V_{eth}/V_d=0.5$.}
\label{figure5}
\end{figure}
 
Assuming an electrostatic field perturbation $\phi=\phi_0\times \rm{exp}(ik \cdot r-i\omega t)$ and a Maxwellian electron distribution function with mean velocity $V_d$ and temperature $T_e$, the dispersion equation can be written in the limit $k_z = 0$ (where $k_z$ is the wavenumber along $B_\perp$) \cite{ducrocq2006}:
 \begin{eqnarray}
 k^2_{xy} \lambda^2_D \left(1-\frac{m}{M} \frac{\omega^2_{pe}}{\omega^2}\right) & = & \sum^{n=\infty}_{n=1} \frac{2{(\omega- k_y V_d)}^2 I_n(b)e^{-b}}{{(\omega-k_y V_d)}^2-{(n\Omega_{ce})}^{2}}\nonumber \\
 & & + (I_0(b)e^{-b} -1)
\end{eqnarray}
where $k_{xy}$ is the wavenumber in the plane perpendicular to $B_\perp$; $\lambda^2_D=k_B T_e/m \omega^2_{pe}$ is the Debye length; $\Omega_{ce}$ is the electron cyclotron frequecy; $\omega_{pe}$ is the electron plasma frequency; $k_y$ is the wavenumber in the $E\times B_\perp$ direction, and $b=k^2_y V^2_{eth}/\Omega^2$. The functions $I_n$ are modified Bessel functions of order n.

In order to find the dependence of $\omega_r$ and $\omega_i$ on the wavenumber $k_y$, we have solved numerically equation (1) by taking a further approximation to neglect the perturbations perpendicular to the drift motion ($k_x=0$, the wavenumber along E) because the electron drift velocity $V_d$ is about an order of magnitude larger than the density gradient drift velocity. Figure~\ref{figure5} shows the frequency and the corresponding growth rate of the unstable modes as a function of $k_y$. Figure~\ref{figure5}a and \ref{figure5}b are the envelopes of the solutions and the first 5 modes, respectively. The values of $f_r$ for the maximum growth rate correspond to frequencies ranging from 90 KHz to 140 KHz for $f_{pe}=21$ MHz, $B_\perp=2$ G, $E=20$ mV/cm and $T_e= 2.7$ K at delay time of about 30 $\mu$s (where $T_e$ is from \cite{fletcher2007}), which agrees with the measured frequencies at $V_d=10^4$ m/s in Fig.~\ref{figure5}. The ratio of the frequency to the drift velocity is about 9-14 $m^{-1}$, in agreement with the measured slope of Fig.~\ref{figure2} (10.2 $m^{-1}$). The maximum of the growth rate is reached for $k_y V_d/\omega_{pe}=1.2$, and the corresponding wavelength is about 0.5 mm, close to the electron gyroradius (about 0.2 mm for $B_\perp=2$ G and $T_e=2.7$ K). We can also see the transitions from stability to instability whenever $k_y V_d /\omega_{pe}$ is close to a cyclotron harmonic $n \Omega_{ce}/\omega_{pe}$. The growth rate reaches a maximum and then decreases sharply between each cyclotron harmonic, and is separated by stable regions. The frequency is several orders of magnitude below the growth rate except in the vicinity of the maximum. 

The frequency roughly linearly depends on the drift velocity (below $k_y V_d /\omega_{pe} =1$), but is independent of the plasma frequency (i.e. plasma density) before the frequency reaches the maximum value (both axis are scaled by electron plasma frequency in Fig.~\ref{figure5}a), which explains the lack of density dependence seen in Fig.~\ref{figure1}. We measure the growth rate by suddenly applying the electric field at different delay times, and find the periodic emission occurs within 1-2 $\mu$s after application of the field, corresponding to a growth rate of $\sim$ 100-150 KHz for $B_\perp=2$ G and $E=20$ mV/cm, which is about the same as the frequency $f_r$. 


 Although we see large scale density structures in the electron images, we do not have a model of the microscopic electron motion that produces the periodic pulses. Note that the applied magnetic field should suppress electron detection, since it is transverse to the detector direction. This is seen in the strong suppression of the prompt electron emission peak. Nevertheless, we detect large pulses of electrons, presumably due to large electron trajectories that extend past the grid (1.5 cm below the plasma) where the large acceleration field can direct the electrons to the detector, even in the presence of the transverse magnetic field.
 
 In summary, we have observed large periodic electron emission and splitting of the electron distribution into two or three lobes from an expanding UCP in the presence of crossed magnetic and electric fields. We identify them as a signature of high-frequency electron drift instability due to the electrons drifting relative to the ions across the magnetic field. We calculate the unstable mode frequencies and growth rates by solving 1D dispersion equation and find good agreement. The large scale changes in the electron spatial distribution remain a mystery, as does the exact mechanism that leads to the emission of electrons. This work shows that UCPs will continue to provide an interesting place to study fundamental plasma physics phenomena.

\begin{acknowledgments}
This work was supported by the National Science Foundation PHY-0714381.
\end{acknowledgments}


\end{document}